\documentstyle{article}
\begin{document} 
\def\1{{\rm 1 \kern -.10cm I \kern .14cm}} 
\def\R{{\rm R \kern -.28cm I \kern .19cm}}

\begin{titlepage} 
\begin{flushright}  LPTHE-ORSAY 96/63 \\UFIFT-HEP-96-25\\
SISSA 145/96/EP \\ hep-ph/9610481 \end{flushright} 
\vskip .8cm 
\centerline{\LARGE{\bf {Quasi-degenerate neutrinos}}} 
\vskip .5cm
\centerline{\LARGE{\bf {from an abelian family symmetry}}}   
\vskip 1.5cm
\centerline{\bf {Pierre Bin\'etruy${}^{a,b)}$\footnote{Visiting Miller 
professor at the Department of Physics of the University of California,
Berkeley}, St\'ephane Lavignac${}^{a)}$\footnote{Supported in 
part by the Human Capital and Mobility
Programme, contract CHRX-CT93-0132.}, 
Serguey Petcov${}^{c)}$\footnote{Also at: Institute of Nuclear Research and Nuclear
Energy, Bulgarian Academy of Sciences, 1784 Sofia, Bulgaria.}
and Pierre Ramond${}^{d)}$\footnote{Supported in part by the 
United States Department of Energy under grant DE-FG05-86-ER40272.}}}   \vskip .5cm
\centerline{\em ${}^{a)}$ Laboratoire de Physique Th\'eorique et Hautes 
Energies\footnote{Laboratoire associ\'e au CNRS-URA-D0063.}}
\centerline{\em Universit\'e Paris-Sud, B\^at. 210,}
\centerline{\em F-91405 Orsay Cedex, France }
\vskip .5cm
\centerline{\em ${}^{b)}$ Department of Physics} 
\centerline{\em and Theoretical Physics Group, Lawrence Berkeley National Laboratory,}
\centerline{\em University of California,}
\centerline{\em Berkeley Ca 94720, USA}
\vskip .5cm
\centerline{\em ${}^{c)}$ Scuola Internazionale Superiore di Studi Avanzati,}
\centerline{\em and Istituto Nazionale di Fisica Nucleare, Sezione
di Trieste} 
\centerline{\em I-34013 Trieste, Italy}
\vskip .5cm
\centerline{\em ${}^{d)}$ Institute for Fundamental Theory,}
\centerline{\em Department of Physics, University of Florida}
\centerline{\em Gainesville FL 32611, USA}

\vskip 2cm
\centerline{\bf {Abstract}}
\vskip 2cm
\indent
We show that  models with an abelian family symmetry which 
accounts for the observed hierarchies of masses and mixings in the quark sector
may also accomodate quasi-degeneracies in the neutrino mass spectrum. 
Such approximate degeneracies 
are, in this context, associated with large mixing angles. The parameters of this class of
models are constrained. We discuss their phenomenological implications for 
present and foreseen neutrino experiments.
\end{titlepage}

%%%%%%%%%%%%%%%%%%%%%%%%%%%%%%%%%%%%%%%%%%%%%%%%%%%%%%%%%%%%%%%%%%%

\section{Introduction}
\indent
Family symmetries which might help to understand the observed
pattern of quark and lepton masses and mixings have received
lately an increasing interest. Even in the simplest example of an
abelian family symmetry \cite{FN,LNS,IR} some general properties
seem to emerge, such as an anomalous behavior \cite{BR} which could be
traced back to an underlying superstring theory. In the sector of
quarks and charged leptons, one is until now limited to
``postdictions'', that is, one tries to explain the already observed
spectrum of masses and Cabibbo-Kobayashi-Maskawa mixing angles. The
neutrino sector on the other hand represents a frontier where one
could in principle make predictions based on the constraints obtained
from the quark and charged lepton sectors. This
would represent a test of these ideas and models.

Of course, the situation is not as straightforward. For example, the
models constructed until now seem to give a hierarchical pattern of
neutrino masses \cite{DLLRS,Elena,GN,Montreal,BLR}, typically masses of the order of a
small parameter $<\theta> /M$ to some power  $n_i$ which decreases
with the family index $i$. It therefore seems that one has to invoke a
non-abelian family symmetry in order to have some sort of degeneracy
between some of the light neutrino masses. We will show in what follows
that this is not necessary and that some abelian family symmetries not
only yield  neutrino mass degeneracies but also predict at which
level the degeneracy is lifted. 

In the context of abelian family symmetries, the presence of
degenerate light neutrinos is   directly related with large mixing
angles. The type of mass spectrum that we consider may thus be of
some use in all scenarios where large mixing angles are needed. One
may mention the large angle branch in the MSW interpretation of the
solar neutrino deficit \cite{MSW} or the angles needed in order to
account for the present atmospheric neutrino data. We will return to
a discussion of the possible phenomenological uses of such models in
Section 4 \footnote{Schemes with three massive  
neutrinos two of 
which or all three are quasi-degenerate in mass, and their phenomenology 
have been discussed in recent articles \cite{2DEGNU,2DEGNUPH} and 
\cite{3DEGNU}, respectively. However, the models considered in refs.
\cite{2DEGNU,3DEGNU} are very different from those we study here.} 
but we would like to keep the discussion as 
general as
possible for the next two sections. In Section 2, we review how to 
use an abelian family symmetry in order to generate mass and
mixing hierarchies. We also describe the class of models we will
consider in the following. Section 3 gives explicit models with two
or three quasi-degenerate neutrinos.

%%%%%%%%%%%%%%%%%%%%%%%%%%%%%%%%%%%%%%%%%%%%%%%%%%%%%%%%%%%%%%%%%%%%%
\vskip 1cm
\section{Neutrino masses and family symmetry}
%%%%%%%%%%%%%%%%%%%%%%%%%%%%%%%%%%%%%%%%%%%%%%%%%%%%%%%%%%%%%%%%%%%%%

The class of models we consider are
extensions of the Minimal Supersymmetric Standard Model (MSSM) with:
\begin{enumerate}
	\item an additional abelian family symmetry, $U(1)_X$. We assume 
that this symmetry is gauged and that its anomalies are compensated
for by the Green-Schwarz mechanism. One could be less restrictive and allow for 
example discrete symmetries. We prefer to stick to continuous gauge symmetries since the
anomaly cancellation provides constraints which seem to go in the right direction for 
phenomenology \cite{BR,BLR},
	\item a MSSM singlet field $\theta$ with $U(1)_X$ charge
$X_{\theta}=-1$.  This singlet is used to break $U(1)_X$ and to generate fermion masses,
 	\item three right-handed neutrinos, $\bar N_i$ ($i$ is a family 
index), in addition to the MSSM spectrum. The light neutrino masses
are then generated by the seesaw mechanism \cite{seesaw}. Of course,
the number of right-handed neutrinos is not constrained by experiment
and could be larger, or smaller, than three. We choose to have one
right-handed neutrino per family mainly for illustration purpose.
\end{enumerate}

In the following, we will concentrate on the lepton sector. We denote 
the lepton fields by $L_i$ (lepton doublets, with their $I_W = + 1/2$
components $\nu_i$), $\bar E_i$ (charged lepton singlets) and $\bar
N_i$ (right-handed neutrinos), and their charges under $U(1)_X$ 
respectively by $l_i$, $e_i$ and $n_i$. We also note $h_u$ and $h_d$
the $U(1)_X$ charges of the two Higgs doublets $H_u$ and $H_d$. 

\subsection{Dirac and Majorana matrices}
\indent

Let us recall briefly how the Dirac (${\cal M}_D$) and Majorana 
(${\cal M}_S$) matrices, which determine the light neutrino spectrum
through the seesaw mechanism, are constrained by the family symmetry.
Each Dirac mass term $L_i \bar N_j H_u$ carries an $X$-charge $p_{ij}
= l_i + n_j + h_u$. If $p_{ij} \neq 0$, the
coupling is forbidden by $U(1)_X$, and the corresponding entry of
${\cal M}_D$ is zero. However, if the excess charge $p_{ij}$ is
positive, one can write non-renormalizable interactions involving the
chiral singlet $\theta$: 
\begin{equation}
  L_i \bar N_j H_u \left( \frac{\theta}{M} \right) ^{p_{ij}}
\end{equation}
where $M$ is a large scale characteristic of the underlying theory 
(typically $M \sim M_{Planck}$ or $M_{GUT}$). When $\theta$ acquires a
vev, $U(1)_X$ is spontaneously broken and effective Dirac masses are
generated: 
\begin{equation}
  ({\cal M}_D)_{ij} \sim v_2 \left( \frac{<\theta>}{M}
\right)^{\:p_{ij}} 
\end{equation} 
where $v_2 = <H_u>$. Since $U(1)_X$
is broken below the scale $M$, $\epsilon \equiv <\theta>/M$ is a small
parameter. Thus the Dirac matrix
obtained has a hierarchical structure, with the order of magnitude of
its entries fixed by their excess charges $p_{ij}$ under $U(1)_X$.

Indeed the same type of analysis has proved to be successful in the
quark sector \cite{IR,JS,BR,DPS,Nir,BLR} where it was shown that
$U(1)_X$ charges can be found which account for the observed
pattern of Dirac masses and mixing angles. Constraints are
numerous and the success of the  procedure is not guaranteed from
the start. Indeed it was shown in \cite{BR,BLR} that, in a large class
of models, the observed masses of the charge $(-1/3)$ quarks and charge
$(-1)$ leptons  constrain the $U(1)_X$ symmetry to be anomalous. This
anomaly must be cancelled through a Green-Schwarz mechanism, which
in turn imposes constraints on the theory and suggests a superstring
origin to the model. Among the constraints is the value of the
Weinberg angle \cite{Ibanez} which turns out to be $\sin^2 \theta_W =
3/8$ and thus surprisingly a successful prediction. If one takes
seriously the superstring nature of the model, then the parameter
$<\theta>/M$ is typically \cite{DSW} of order $10^{-2}$ to $10^{-1}$,
that is of the order of the sine of the Cabibbo angle, the basic mixing
angle in the CKM matrix. 

The entries of the Majorana matrix ${\cal M}_S$ are generated in the 
same way, with non-renormalizable interactions of the form:
\footnote{The two mass parameters which appear here and are, for 
simplicity, both denoted by $M$, may be different \cite{Montreal,BLR}. 
We will not discuss here in more details the nature of these scales and 
how they might give rise after seesaw to the right scale of neutrino mass.}
\begin{eqnarray}
  M \bar N_i \bar N_j \left( \frac{\theta}{M} \right) ^{q_{ij}}
\end{eqnarray}
giving rise to effective Majorana masses
\begin{equation}
  ({\cal M}_S)_{ij} \sim M \left( \frac{<\theta>}{M}
\right)^{\:q_{ij}} 
\end{equation}
provided that $q_{ij} = n_i + n_j$ is a positive integer (otherwise 
$({\cal M}_S)_{ij} = 0$).

The structure of the light neutrino mass matrix 
\begin{equation}
{\cal M}_{\nu}  = - {\cal M}_D {\cal M}^{-1}_S {\cal M}^T_D 
\end{equation}
is therefore fixed by the charges of the leptons under $U(1)_X$. Note,
however, that {\em each of the entries is determined only up to an
arbitrary factor of order one by the family symmetry}. Irrespective
of these factors, the Majorana mass matrix ${\cal M}_S$ and therefore
the light neutrino mass matrix ${\cal M}_\nu$ are automatically
symmetric.
The presence of these unknown factors of order one is certainly the 
most unwelcome feature of this type of approach when we come to detailed 
predictions since we will then take the parameter $\epsilon$ to be 
$\sin \theta_c \sim 0.22$, which is not such a small number. The
only way to avoid this type of problem would be to go to a specific model. 
Since we want to advertize the general features of such a class of models, 
we will refrain from doing so. But, keeping this in mind, all the constraints 
that we will obtain below on the $U(1)_X$ lepton charges will be
understood up to one unit.

\subsection{Light neutrino spectrum}
\indent

If all entries of ${\cal M}_D$ and ${\cal M}_S$ are nonzero 
(i.e. $p_{ij}$, $q_{ij} \geq 0$), one obtains:
\begin{equation}
  ({\cal M}_{\nu})_{ij} \sim {v_2^{\: 2} \over M} \; \epsilon^{\: l_i+l_j+2h_u}
\label{eq:Mnu}
\end{equation}
which leads to the following light neutrino masses and mixings:
\begin{eqnarray}
  m_{\nu_i} \sim {v_2^{\: 2} \over M} \; \epsilon^{\: 2l_i+2h_u}  &  &
  (R_{\nu})_{ij} \sim \epsilon^{\: |l_i-l_j|}
\label{eq:spectrum}
\end{eqnarray}
Here $R_{\nu}$ is the matrix that diagonalizes ${\cal M}_{\nu}$:  
${\cal M}_{\nu} = R_\nu Diag(m_{\nu_1}, m_{\nu_2}, m_{\nu_3})  R_{\nu}^T $. 
The physical mixing angles are given by the lepton mixing matrix 
 $U = R^L_e R_\nu$ \footnote{The neutrino mixing was first introduced 
in ref. \cite{Pont57}. The mixing of neutrinos having different flavour
was considered first in ref. \cite{MNS62}, while a Majorana 
mass term for the left-handed flavour neutrinos was discussed as a possible 
source of lepton mixing first in ref. \cite{GP69}.}, 
which also includes a contribution from 
the charged lepton sector: $R^L_e {\cal M}_e R^{R \: \dagger}_e = Diag(m_1, m_2, m_3)$. 
When the excess charges of the charged lepton Yukawa couplings $L_i \bar E_j H_d$ 
are positive, $R^L_e$ has the same structure as $R_{\nu}$, and $U$ is given by: 
\begin{equation}
  U_{ij} = (R^L_e R_\nu)_{ij} \sim \epsilon^{\: |l_i-l_j|}
\end{equation}
Then one has the relation between neutrino mass ratios and mixing 
angles:
\begin{eqnarray}
  U^2_{ij} \sim {m_{\nu_i} \over m_{\nu_j}}  &  &  m_{\nu_i} < m_{\nu_j}
\end{eqnarray}
A general feature of such models is that the neutrino mass spectrum is 
naturally hierarchical, with small mixing angles. Mass degeneracies
occur when some lepton charges are equal (for example, if $l_2 = l_3$,
one obtains $m_{\nu_2} \sim m_{\nu_3}$). Unfortunately, the model is
less predictive in this case. Indeed, because of  the presence of 
arbitrary factors of order one in each entry of ${\cal M}_{\nu}$, the
squared mass difference between almost degenerate neutrinos cannot be
related to the parameters of the family symmetry\footnote{ Also, in
presence of degeneracies, these factors   could upset the formulae
(\ref{eq:spectrum}).}. An accurate mass degeneracy would then require
fine-tuning. This problem could be solved by going to a larger
non-abelian family symmetry, which would fix the factors of order one.
We will however follow a different path here: keep abelian family
symmetries but assume zeroth order (in $\epsilon$) relations among
the Yukawa couplings which ensure degeneracies. 

Indeed, if ${\cal M}_D$ and ${\cal M}_S$ contain some zeros, the
above  results can be modified. One can show that, if there are
simultaneous and correlated zeros in ${\cal M}_D$ and ${\cal M}_S$,
some entries of ${\cal M}_{\nu}$ become zero, the other entries being
still given by (\ref{eq:Mnu}). Of course, the neutrino spectrum then
deviates from (\ref{eq:spectrum}). This raises the hope that the
presence of correlated zeroes in ${\cal M}_D$ and ${\cal M}_S$ can
explain mass degeneracies and large mixing angles.

%%%%%%%%%%%%%%%%%%%%%%%%%%%%%%%%%%%%%%%%%%%%%%%%%%%%%%%%%%%%%%%%%%%%%%
\vskip 1cm
\section{Models of quasi-degenerate neutrinos}
%%%%%%%%%%%%%%%%%%%%%%%%%%%%%%%%%%%%%%%%%%%%%%%%%%%%%%%%%%%%%%%%%%%%%%
\indent

In order to find abelian family symmetries leading to quasi-degenerate 
neutrinos, we proceed in the following way. We start from a very
simple pattern for the light neutrino matrix ${\cal M}_\nu$ (here we
assume that the seesaw mechanism has already been performed), with a
little number of nonzero entries and two or three exactly degenerate
eigenvalues. Then, assuming a $U(1)_X$ symmetry, we identify the most
general $X$-charges compatible with this pattern. The
breaking of $U(1)_X$ fills in the zero entries in ${\cal M}_\nu$ with
powers of the small parameter $\epsilon$. This lifts the degeneracy
between the neutrinos and allows us to relate the squared mass
differences to the $X$-charges. Note that this method
is similar to the one used in the quark sector where,
in order to account for the strong hierarchy of the quark mass
spectrum, one starts from an up quark matrix with all entries zero
except the (3,3) entry.

\subsection{Patterns of neutrino mass matrices with degenerate eigenvalues}
\indent

Consider the following three symmetric matrices \cite{P,WW}:
\begin{eqnarray}
  (i) \; \; \; {\cal M}_{\nu} = \left( \begin{array}{ccc}
    0 & 0 & 0 \\
    0 & 0 & a \\
    0 & a & 0   \end{array}  \right)  &  &
  (ii) \; \; \; {\cal M}_{\nu} = \left( \begin{array}{ccc}
    0 & 0 & b \\
    0 & 0 & a \\
    b & a & 0   \end{array}  \right)  \nonumber
\end{eqnarray}
\begin{equation}
  (iii) \; \; \; {\cal M}_{\nu} = \left( \begin{array}{ccc}
    b & 0 & 0 \\
    0 & 0 & a \\
    0 & a & 0   \end{array}  \right)
\end{equation}
where $a$ and $b$ are arbitrary numbers of order one. These matrices 
have two degenerate eigenvalues\footnote{The negative values in the
diagonal mass matrix $D_{\nu}$ imply that the corresponding
Majorana neutrinos with positive definite mass have a
CP-parity equal to $(-i)$, while the CP-parity of the other
definite mass Majorana neutrinos is $(+i)$ \cite{CP}. Two
mass-degenerate Majorana neutrinos having opposite CP-parities are
equivalent to a Dirac neutrino.} and 
their corresponding diagonalizing matrices $R_{\nu}$ have  
at least one large mixing angle.
Explicitly: 
\begin{eqnarray}
  (i)  &  D_{\nu} = \left( \begin{array}{ccc}
    0 & 0 & 0 \\
    0 & -a & 0 \\
    0 & 0 & a   \end{array}  \right)  &
  R_{\nu} = \left( \begin{array}{ccc}
    1 & 0 & 0 \\
    0 & {1 \over \sqrt{2}} & {1 \over \sqrt{2}} \\
    0 & -{1 \over \sqrt{2}} & {1 \over \sqrt{2}}   \end{array}  \right)
\end{eqnarray}
\begin{eqnarray}
  (ii)  &  D_{\nu} = \left( \begin{array}{ccc}
    0 & 0 & 0 \\
    0 & -\sqrt{a^2+b^2} & 0 \\
    0 & 0 & \sqrt{a^2+b^2}   \end{array}  \right)  &  \nonumber
\end{eqnarray}
\begin{equation}
  R_{\nu} = {1 \over \sqrt{2(a^2+b^2)}} \left( \begin{array}{ccc}
    -\sqrt{2} a & -b & b \\
   \sqrt{2} b  & -a & a \\
    0 &  \sqrt{a^2+b^2} & \sqrt{a^2+b^2}   \end{array}  \right)
\end{equation}
\begin{eqnarray}
  (iii)  &  D_{\nu} = \left( \begin{array}{ccc}
    b & 0 & 0 \\
    0 & -a & 0 \\
    0 & 0 & a   \end{array}  \right)  &
  R_{\nu} = \left( \begin{array}{ccc}
    1 & 0 & 0 \\
    0 & {1 \over \sqrt{2}} & {1 \over \sqrt{2}} \\
    0 & - {1 \over \sqrt{2}} & {1 \over \sqrt{2}}   \end{array}  
\right)
\end{eqnarray}
where $D_\nu = Diag(m_{\nu_1}, m_{\nu_2}, m_{\nu_3}) 
= R^T_\nu {\cal M}_\nu R_\nu$. It can be easily seen that the previous
three patterns are, up to permutations in family space, the only
matrices with arbitrary nonzero
numbers having at least two degenerate eigenvalues (up to a sign). Three degenerate
eigenvalues cannot be obtained with arbitrary entries. One needs for
example to impose $a=b$ in pattern (iii), which requires an additional
symmetry.

It is not difficult to find the most general $X$-charge reproducing 
pattern (i), (ii) or (iii) when unbroken. The charge operator acting
on lepton doublets, $X_L$, can be identified with a generalized 
Zeldovich-Konopinsky-Mahmoud \cite{ZKM}  combination of
lepton numbers \cite{P}: patterns (ii) and (iii) correspond respectively to $X_L =
l \, (L_e + L_\mu - L_\tau)$ and $X_L = l \, (L_\mu - L_\tau)$, where
$l$ is a rational number. Pattern (i) corresponds to $X_L = l' L_e + l
\, (L_\mu - L_\tau)$ with $l' \neq -l, l$, which means that $L_e$ and
$L_\mu - L_\tau$ are separately conserved. 

The breaking of the family symmetry fills in the zero entries with powers 
of the small parameter $\epsilon$, namely $({\cal M}_{\nu})_{ij} \sim 
\epsilon^{\: l_i+l_j}$ when $p_{ij}, q_{ij} \geq 0$. If all powers are 
positive, this slightly modifies the previous results. 
In particular, the degeneracy between $m_{\nu_2}$ and 
$m_{\nu_3}$ is broken\footnote{ Thus transforming the corresponding Dirac 
neutrino into a pseudo-Dirac one.}, and $m_{\nu_1}$ becomes nonzero in cases 
(i) and (ii). If, on the other hand,
some powers are negative, the previous patterns are destabilized. 
This happens in fact in all three cases. Assuming $l \geq 0$, one obtains 
$({\cal M}_\nu)_{33} \sim \epsilon^{\: -2l} \gg 1$, thus the (3,3) entry is 
the dominant one after breaking of the
family symmetry, and the mass spectrum becomes hierarchical. One can easily 
check that the only way to avoid such effect is to assume the 
presence of simultaneous and correlated zeroes in ${\cal M}_D$ and 
${\cal M}_S$, which forces $({\cal M}_\nu)_{33}$ to vanish.

\subsection{Explicit models with 2 degenerate neutrinos}
\indent

In this section, we study an explicit model with 2 degenerate 
neutrinos based on the pattern (i). We will briefly comment on models
based on patterns (ii) and (iii) at the end of the section.

In order to reproduce pattern (i), we choose the following assignment 
of lepton charges:
\begin{eqnarray}
  \left\{ \begin{array}{l}
	l_1 = l' \\ l_2 = - l_3 = l \\ 0 < l < l'  \end{array}  \right.
  &  \mbox{and}  &  0 \leq n_1 < (-n_3) \leq l \leq n_2
\label{eq:assignment}
\end{eqnarray}
and, for simplicity, we  assume $h_u=h_d=0$.

The model thus contains five parameters $l$, $l'$, $n_1$, $n_2$ and 
$n_3$. With this charge assignment, the Dirac and Majorana matrices
have correlated zeroes: \begin{eqnarray}
  {\cal M}_D \sim \left( \begin{array}{ccc}
    \epsilon^{\: l'+n_1} & \epsilon^{\: l'+n_2} & \epsilon^{\: l'+n_3} \\
    \epsilon^{\: l+n_1} & \epsilon^{\: l+n_2} & \epsilon^{\: l+n_3} \\
    0 & \epsilon^{\: -l+n_2} & 0   \end{array}  \right)  \\
   {\cal M}_S \sim \left( \begin{array}{ccc}
    \epsilon^{\: 2n_1} & \epsilon^{\: n_1+n_2} & 0 \\
    \epsilon^{\: n_1+n_2} & \epsilon^{\: 2n_2} & \epsilon^{\: n_2+n_3} \\
    0 & \epsilon^{\: n_2+n_3} & 0   \end{array}  \right)
\end{eqnarray}
which results in the following structure for the light neutrino matrix:
\begin{equation}
  {\cal M}_\nu \sim \left( \begin{array}{ccc}
     \: \epsilon^{\: 2l'} &  \: \epsilon^{\: l'+l} &
	 \: \epsilon^{\: l'-l} \\
     \: \epsilon^{\: l'+l} &  \: \epsilon^{\: 2l} & 1 \\
     \: \epsilon^{\: l'-l} & 1 & 0   \end{array}  \right)
\end{equation}
where we have left aside the overall mass scale set by the seesaw mechanism. 
Note that, due to the simultaneous presence of zeroes in ${\cal M}_D$
and ${\cal M}_S$, the  structure of ${\cal M}_\nu$ follows pattern
(i) to zeroth order in the small parameter $\epsilon$. The main effect
of the family symmetry breaking is  to break slightly the
degeneracy between $m_{\nu_2}$ and $m_{\nu_3}$. 
We thus obtain a pattern with two almost degenerate neutrinos 
\begin{equation}
m_{\nu_1} \sim \epsilon^{2 l'} \ll - \: m_{\nu_2} \simeq m_{\nu_3},
\end{equation}
their squared mass difference being determined by the family symmetry: 
\begin{equation}
  \Delta m^2_{32} \; \sim \;  \epsilon^{ 2l}
\end{equation}

Actually, one should take into account the presence of nondiagonal 
kinetic terms in the K\"{a}hler potential which are
allowed by the symmetry \cite{LNS,DPS,BLR}: 
\begin{equation}
\sum_{i,j}  L^{\dagger}_i L_j \left[ H(l_i-l_j) \left(
\frac{\theta^{\dagger}}{M} \right) ^{l_i-l_j} + H(l_j-l_i) \left(
\frac{\theta}{M} \right) ^{l_j-l_i} \right].
\end{equation} 
where $H$ is the Heaviside function ($H(x)=x$ if $x
\geq 0$,  $H(x)=0$ otherwise). The lepton fields have to be redefined
in order to bring the K\"{a}hler potential into its canonical form:
\begin{equation}
  L_i \rightarrow (W_L)_{ij} L_j
\end{equation}
where the matrix elements of $W_L$ are constrained in order of 
magnitude by the lepton charges:
\begin{equation}
  (W_L)_{ij} \sim \epsilon^{\: |l_i-l_j|}\; .
\end{equation}
The light neutrino mass  matrix, expressed in the new basis, is then:
\begin{equation}
  \widehat{{\cal M}}_\nu = W^T_L {\cal M}_\nu W_L
\end{equation}
The effect of this redefinition is to fill in the zero entries in 
${\cal M}_\nu$, the other entries remaining of the same order of
magnitude. For the model under consideration: 
\begin{equation}
  \widehat{\cal M}_\nu \simeq \left( \begin{array}{ccc}
    b \: \epsilon^{\: 2l'} & c \: \epsilon^{\: l'+l} &
	e \: \epsilon^{\: l'-l} \\
    c \: \epsilon^{\: l'+l} & d \: \epsilon^{\: 2l} &
	a \\
    e \: \epsilon^{\: l'-l} & a &
	f \: \epsilon^{\: 2l}   \end{array}  \right) \label{eq:hatMnu}
\end{equation}
where only the dominant term in each entry is given\footnote{The 
other terms introduced by the lepton field redefinition have been
taken into account in the diagonalization of $\widehat{{\cal M}}_\nu$.}.
We have restored in (\ref{eq:hatMnu}) arbitrary coefficients of 
order one $a$, $b$, $c$, $d$, $e$ and $f$.

The eigenvalues are not significantly modified by the  filling of
previously vanishing entries.  New subleading terms (which we will
omit for the sake of simplicity) appear in their expressions, but
their orders of magnitude remain the same: 
\begin{equation}
  \left \{  \begin{array}{lll}
	m_{\nu_1} & \simeq &   
	  \frac{a^2b+de^2-2ace}{a^2} \; \epsilon^{\: 2l'}  \\
	-m_{\nu_2} & \simeq m_{\nu_3} & \simeq \; a  \\
	m_{\nu_2} + m_{\nu_3} & \simeq & (d+f) \: \epsilon^{\: 2l} 
\end{array} \right. \end{equation}
Thus, the mass splitting between $m_{\nu_2}$ and 
$m_{\nu_3}$  remains of the same order:
\begin{equation}
  \Delta m^2_{32} \; \simeq \; 2 \: a (d+f) \: \epsilon^{\: 2l}
\end{equation}
The other squared mass differences are:
\begin{equation}
  \left \{  \begin{array}{lll}
	\Delta m^2_{21} & \simeq & [a^{\: 2} \: + \: \ldots ]
	  - \: [a (d+f) \: \epsilon^{\: 2l} \: + \: \ldots ] \\
	\Delta m^2_{31} & \simeq & [a^{\: 2} \: + \: \ldots ]
	  + \: [a (d+f) \: \epsilon^{\: 2l} \: + \: \ldots ] 
\end{array} \right.
\end{equation}
The diagonalization matrix $R_\nu$ is:
\begin{equation}
  R_\nu \; \simeq \; \left( \begin{array}{ccc}
	1 & \frac{e}{\sqrt{2} a} \: \epsilon^{\: l'-l} & \frac{e}{\sqrt{2} a} \: \epsilon^{\: l'-l} \\
\: - \frac{e}{a} \: \epsilon^{\: l'-l}	&
\frac{1}{\sqrt{2}} & \frac{1}{\sqrt{2}}  \\
\frac{de - ac}{a^2} \: \epsilon^{\: l'+l}	 &
-\frac{1}{\sqrt{2}} & \; \; \: \frac{1}{\sqrt{2}}
  \end{array}  \right)
\end{equation}
where only the dominant term in each entry is given. As expected, the 
mixing between the two degenerate neutrinos remains large after
breaking of $U(1)_X$ and zero filling.

Since the physical mixing angles, i.e. the mixing angles which are 
relevant for neutrino oscillations, are given by the lepton mixing
matrix $U = R^L_e R_\nu$, we must study the charged
lepton sector too. Like for the neutrinos, the order of magnitude of
the charged lepton Yukawa couplings $L_i \bar E_j H_d$ are determined
by the charges of the fields $L_i$ and $\bar E_i$ under $U(1)_X$:
\begin{equation}
  ({\cal M}_e)_{ij} \sim v_1 \left( \frac{<\theta>}{M} 
\right) ^{\:n_{ij}}
\end{equation}
where $v_1 = <H_d>$ and  $n_{ij} = l_i+e_j$ (we have assumed 
$h_d=0$). The charges of the
lepton doublets have already been fixed. Obviously, one cannot choose
the same type of charge assignment for the charged lepton singlets
$\bar E_i$, otherwise there would be two degenerate charged
leptons\footnote{If $e_2 = - e_3 = l$, the charged lepton matrix
${\cal M}_e$ has its dominant entries in positions (2,3) and (3,2).}.
The most natural structure for ${\cal M}_e$ is a hierarchical
structure, as for the quark mass matrices, with dominant entry in
position (3,3). When all $n_{ij}$ are positive, the charged lepton
mass matrix has the following form: \begin{equation}
  {\cal M}_e \sim \left( \begin{array}{lll}
    \epsilon^{\: l'+e_1} & \epsilon^{\: l'+e_2} & \epsilon^{\: l'+e_3} \\
    \epsilon^{\: l+e_1} & \epsilon^{\: l+e_2} & \epsilon^{\: l+e_3} \\
    \epsilon^{\: -l+e_1} & \epsilon^{\: -l+e_2} & \epsilon^{\: -l+e_3}
  \end{array}  \right) \label{eq:Me}
\end{equation}
When some $n_{ij}$ are negative, the corresponding entries of 
${\cal M}_e$ are modified (after zero filling). ${\cal M}_e$ is in
general not hermitian, so it is diagonalized by two unitary matrices: 
$Diag(m_1, m_2, m_3) = R^L_e {\cal M}_e R^{R \: \dagger}_e$. Assuming
simply $({\cal M}_e)_{33} \geq ({\cal M}_e)_{ij}$ and $({\cal
M}_e)_{33} \gg ({\cal M}_e)_{ij}$ for $i,j=1,2$ (in order to obtain a
hierarchical mass spectrum), one can show that the diagonalization
matrix $R^L_e$, which enters the lepton mixing matrix, is given by:
\begin{equation}
  R^L_e \; \sim \; \left( \begin{array}{ccc}
	1 & \epsilon^{\: l'-l} & \epsilon^{\: l'+l}  \\
	\epsilon^{\: l'-l} & 1 & \epsilon^{\: 2l}  \\
	\epsilon^{\: l'+l} & \epsilon^{\: 2l} & 1
  \end{array}  \right)
\end{equation}
except in the very particular case where $e_1 \geq e_3 > e_2$ 
and $l'-l \geq e_3-e_2 > 2l$. The lepton mixing matrix is then:
\begin{equation}
  U \; \simeq \; \left( \begin{array}{ccc}
    \; \; \; 1 & \; \; \; A \: \epsilon^{\: l'-l} &
	A \: \epsilon^{\: l'-l}  \\
    - \sqrt{2} \, A^{*} \: \epsilon^{\: l'-l} & \; \; \:
	{1 \over \sqrt{2}} & {1 \over \sqrt{2}}  \\
    \; \; \; B \: \epsilon^{\: l'+l} & - {1 \over \sqrt{2}} &
	{1 \over \sqrt{2}}
  \end{array}  \right)
  \label{eq:mixing}
\end{equation}
where $A$ and $B$ are functions of the arbitrary order one factors entering ${\cal M}_\nu$ 
and ${\cal M}_e$. Unless unnatural cancellations occur, $A$ and $B$ should not be much different 
from 1, say $0.2 < A, B < 5$.

Note that due to the  smallness of the $R^L_e$ entries, $U$ has the 
same structure as $R_\nu$. As for the charged lepton masses, their
experimental values constrain not only the singlet lepton charges
$e_1$, $e_2$ and $e_3$, but also the doublet charges $l$ and $l'$.

Let us now briefly comment on models based on patterns (ii) and (iii). 
Pattern (ii) can be obtained by setting $l'=l$ in the charge
assignment (\ref{eq:assignment}). Thus $l_1 = l_2 = - l_3 = l$, and
${\cal M}_\nu$ depends on one charge parameter only: 
\begin{equation}
  {\cal M}_\nu = \left( \begin{array}{ccc}
    b \: \epsilon^{\: 2l} & c \: \epsilon^{\: 2l} & e \\
    c \: \epsilon^{\: 2l} & d \: \epsilon^{\: 2l} & a \\
    e & a & 0   \end{array}  \right)
\end{equation}
where $a$, $b$, $c$, $d$, $e$ are arbitrary coefficients of order one. 
There are still two degenerate neutrinos ($m_{\nu_1} \ll - \: m_{\nu_2} \simeq m_{\nu_3}$) 
with a squared mass difference of order $\epsilon^{\: 2l}$, and a light neutrino with mass 
$m_{\nu_1} \sim \epsilon^{\: 2l}$. The diagonalization matrix $R_\nu$ has only one small entry:
\begin{equation}
  R_\nu \; \simeq \; \left( \begin{array}{ccc}
	- \frac{a}{\sqrt{a^2+e^2}} & - \; \; \: \frac{e}{\sqrt{2(a^2+e^2)}} 
&  \; \; \: \frac{e}{\sqrt{2(a^2+e^2)}} \\
	\frac{e}{\sqrt{a^2+e^2}} &
- \frac{a}{\sqrt{2(a^2+e^2)}} & \; \; \: \frac{a}{\sqrt{2(a^2+e^2)}}  \\ 
\frac{ae(b-d)+c(a^2-e^2)}{(a^2+e^2)^{3/2}} \: \epsilon^{\: 2l}
	 & \frac{1}{\sqrt{2}}
 & \frac{1}{\sqrt{2}}
  \end{array}  \right)
\end{equation}
Pattern (iii) can be obtained by setting $l'=0$ in the charge 
assignment (\ref{eq:assignment}). Thus $l_1 = 0$ and $l_2 = - l_3 =
l$, and again ${\cal M}_\nu$ depends on one charge parameter only:
\begin{equation}
  {\cal M}_\nu = \left( \begin{array}{ccc}
    b & c \: \epsilon^{\: l} & 0 \\
     c \: \epsilon^{\: l} & d \: \epsilon^{\: 2l} & a \\
    0 & a & 0   \end{array}  \right)
\end{equation}
where $a$, $b$, $c$, $d$ are arbitrary coefficients of order one. 
The three masses lie in the same range, two of them being
quasi-degenerate ($m_{\nu_1} \sim - \: m_{\nu_2} \simeq m_{\nu_3}$),
with a squared mass difference of order $\epsilon^{\: 2l}$. The
diagonalization matrix is: \begin{equation}
  R_\nu \; \simeq \; \left( \begin{array}{ccc}
	1 & - \frac{c}{\sqrt{2} (a+b)} \: \epsilon^{\: l} &
  \; \; \: \frac{c}{\sqrt{2} (a-b)} \: \epsilon^{\: l} \\
\: - \frac{bc}{a^2-b^2} \: \epsilon^{\: l}  &
\frac{1}{\sqrt{2}} &  \frac{1}{\sqrt{2}}  \\
\: - \frac{ac}{a^2-b^2} \: \epsilon^{\: l}	 &
-\frac{1}{\sqrt{2}} & \; \; \: \frac{1}{\sqrt{2}}
  \end{array}  \right)
\end{equation}

\subsection{Case of 3 degenerate neutrinos}
\indent

As mentioned above, pattern (iii) with $a=b$ leads to three 
degenerate eigenvalues. Unfortunately, this cannot be obtained from a
U(1) family symmetry alone. One needs an additional symmetry to
explain why $a=b$. Let us assume that such a symmetry exists. With the
charge assignment of previous section ($l_1 = 0$ and $l_2 = - l_3 =
l$), we get: 
\begin{equation}
  {\cal M}_\nu = \left( \begin{array}{ccc}
    a_0 & c_0 \: \epsilon^{\: l} & 0 \\
     c_0 \: \epsilon^{\: l} & d_0 \: \epsilon^{\: 2l} & a_0 \\
    0 & a_0 & 0   \end{array}  \right)
\end{equation}
Taking into account the presence in the Lagrangian of the theory 
of nondiagonal kinetic terms allowed by the symmetry we obtain:
\begin{equation}
  \widehat{\cal M}_\nu \simeq \left( \begin{array}{ccc}
    a & c \: \epsilon^{\: l} &
	e \: \epsilon^{\: l} \\
    c \: \epsilon^{\: l} & d \: \epsilon^{\: 2l} &
	a \\
    e \: \epsilon^{\: l} & a &
	f \: \epsilon^{\: 2l}   \end{array}  \right)
\end{equation}

The calculation of the eigenvalues up to order $\epsilon^{\: l}$ gives:
\begin{equation}
  \left \{  \begin{array}{llrl}
	m_{\nu_1} & \simeq & a & - \: \frac{c + e}{\sqrt{2}} \: \epsilon^{\: l}  \\
	m_{\nu_2} & \simeq & - \: a &  \\
	m_{\nu_3} & \simeq & a & + \: \frac{c + e}{\sqrt{2}} \: \epsilon^{\: l}
  \end{array} \right.
\end{equation}
We obtain three almost degenerate neutrinos ($m_{\nu_1} \simeq 
- \: m_{\nu_2} \simeq m_{\nu_3}$), with all squared mass differences
of the same order 
\footnote{Obviously, 
the same pattern of neutrino masses (i.e., three quasi-degnerate neutrinos)
will arise also if instead of the charge $X_{L} = l\, (L_{\mu} - L_{\tau})$ the charge 
$X_{L} = l\, (L_{\mu} - L_{e})$ is conserved before the spontaneous breaking of 
the $U(1)_{X}$ symmetry and the two different large entries in ${\cal M}_\nu$ are equal.}: 
\begin{equation}
  \left \{  \begin{array}{llr}
	\Delta m^2_{21} & \simeq & \sqrt{2} \: a(c + e)\: \epsilon^{\: l}  \\
	\Delta m^2_{31} & \simeq & 2 \sqrt{2} \: a(c + e) \: \epsilon^{\: l}  \\
	\Delta m^2_{32} & \simeq & \sqrt{2} \: a(c + e)\: \epsilon^{\: l}
  \end{array} \right.
\end{equation}
The diagonalization matrix and correspondingly the lepton mixing matrix 
have only one small entry
\begin{equation}
  R_\nu \; \simeq \; \left( \begin{array}{ccc}
	\frac{1}{\sqrt{2}} & - \frac{c - e}{2 \sqrt{2} a} \: \epsilon^{\: l}
 & \frac{1}{\sqrt{2}}  \\
	- \frac{1}{2}   &
	  \frac{1}{\sqrt{2}} &  \frac{1}{2}  \\
	- \frac{1}{2} & \; \; \: - \frac{1}{\sqrt{2}} & \; \; \: \frac{1}{2}
  \end{array}  \right)
\end{equation}
 
\begin{equation}
  U \; \simeq \; \left( \begin{array}{ccc}
	\frac{1}{\sqrt{2}} & A \: \epsilon^{\: l}
 & \frac{1}{\sqrt{2}}  \\
	- \frac{1}{2}   &
	  \frac{1}{\sqrt{2}} &  \frac{1}{2}  \\
	- \frac{1}{2} & \; \; \: - \frac{1}{\sqrt{2}} & \; \; \: \frac{1}{2}
  \end{array}  \right)
\end{equation}
\noindent The corrections in the entries of order one in the matrices $R_{\nu}$ and
$U$ are of the order of $\epsilon^{\: l}$. 

\section{Phenomenological analysis}
\indent

Experimental data yield strong indications of neutrino oscillations. Most convincing 
is the solar neutrino deficit \cite{SNP} and its MSW interpretation 
\cite{MSW}. The solar neutrino problem admits also a vacuum oscillation 
solution \cite{VOS}. Present data implies (see, e.g., \cite{KP}) two 
possible sets of fundamental parameters in the MSW case: 
the adiabatic branch requires large mixing, and the non-adiabatic 
branch requires small mixing of the electron neutrino with 
another specie of neutrino. 
The vacuum oscillation interpretation of the data is possible only if the 
indicated mixing is large \cite{KP}.  
Several experiments (Super-Kamiokande, SNO, BOREXINO, ICARUS, HELLAZ) should in 
the future allow to distinguish
between these solutions \footnote{The Super-Kamiokande experiment began operation
on April 1, 1996, and the completion of SNO is scheduled for the
beginning of 1997.}. Measurements of the fluxes of electron and muon neutrinos
resulting from cosmic ray interactions 
in the Earth atmosphere indicate a deficiency of muon neutrinos 
\cite{KAM,ANP1,ANP2}. This 
anomaly can be caused by oscillations of the atmospheric muon neutrinos
into another specie of neutrino with a large mixing angle \cite{KAM,ANP2}. 
The region of the parameter space the 
neutrino oscillation solution of the atmospheric neutrino problem implies 
will soon be tested in accelerator experiments. Lastly, the LSND 
group has 
reported \cite{LSND} evidence for $\bar \nu_\mu - \bar \nu_e$ oscillations 
with a small mixing angle. Of course, there remains to be seen 
how all these experimental results will stand the test of time. 

There is also some cosmological rationale for one or several neutrinos with 
masses of a few $eV$, to add the right amount of hot dark matter to cold dark 
matter in order to reproduce structure formation in the Universe 
\cite{Primack}.

As stressed in the introduction, if one wants to explain all mass hierarchies 
with abelian family symmetries, there is a one-to-one correspondance 
between large mixing angles and mass degeneracies. Indeed, unless some 
fine-tuning occurs in the mass matrix
${\cal M}_\nu$, quasi-degenerate neutrinos automatically have a large 
mixing, while neutrinos well separated in mass have a small mixing. 
This implies that the MSW adiabatic solution (AS) or 
the vacuum oscillation solution (resp. 
the interpretation of the atmospheric neutrino data in
terms of oscillations) requires the electron neutrino 
(resp. the muon neutrino) to be degenerate in mass 
\footnote{More precisely, we mean the massive neutrino whose state 
vector is the dominant component in the electron neutrino state vector
when expressed (with the help of the lepton mixing matrix U) 
as linear combination of the state vectors of the neutrinos $\nu_{i}$, i=1,2,3,
with definite mass, etc.}
with another kind of neutrino,  
while the MSW non-adiabatic solution (NAS) as well as the oscillation 
explanation of the LSND result involve neutrinos with a 
substantial mass splitting. 
The vacuum oscillation solution of the solar neutrino 
problem can be realized only if the massive neutrinos are highly degenerate 
in mass ($\Delta m^2 \simeq (10^{-10} - 5.10^{-12})~{\rm eV^2}$ 
\cite{KP}), which in turn requires huge and seemingly unrealistic 
values of the charge $l$. For this reason we shall not consider it further. 

  Keeping the above in mind, one can show that only six neutrino mass 
patterns are 
compatible with at least two of the three possible ``unconventional'' interpretations
of the three indicated experimental results, namely 
the MSW transitions of solar neutrinos ($\nu_{\odot}$) , 
the atmospheric neutrino oscillations and the 
small angle $\bar \nu_\mu - \bar \nu_e$ oscillations. 
These mass patterns are:

\begin{enumerate}
\item  $m_{\nu_{\tau(e)}} \ll m_{\nu_{e(\tau)}} \ll m_{\nu_\mu}$ ($\nu_{\odot}-$problem
(MSW NAS) and LSND effect)
\item  $m_{\nu_e} \simeq m_{\nu_\tau} \ll m_{\nu_\mu}$ ($\nu_{\odot}-$problem
(MSW AS) and LSND effect)
\item  $m_{\nu_\mu} \ll m_{\nu_e} \simeq m_{\nu_\tau}$ ($\nu_{\odot}-$problem
(MSW AS) and LSND effect)
\item  $m_{\nu_e} \ll m_{\nu_\mu} \simeq m_{\nu_\tau}$  (atmospheric$-\nu$ problem 
and LSND effect)
\item  $m_{\nu_e} \simeq m_{\nu_\mu} \simeq m_{\nu_\tau}$ 
($\nu_{\odot}-$problem
(MSW AS) and atmospheric$-\nu$ problem)
\item  $m_{\nu_\tau}, m_{\nu_\mu} \ll m_{\nu_e}$  (atmospheric$-\nu$ problem and LSND effect)
\end{enumerate}
Patterns $3$, $4$ and $5$ can account for the hot dark matter of the Universe, 
while in patterns $1$,$2$ and $6$, the heavier neutrino may be too light for this purpose. 
Note that it is not possible to accommodate 
all present data simultaneously with only three species of light neutrinos,
if one ever wanted to. 

We will adopt in this section the following strategy. We start with either of
the two problems whose neutrino physics solutions involve large lepton 
mixing angles (atmospheric neutrino anomaly
or the solar neutrino problem (MSW adiabatic solution)) and try to 
accomodate them in the framework
of the models presented earlier. This in turn constrains the lepton charges.
A further important constraint arises from the charged lepton sector where one 
wants to reproduce the observed mass hierarchy. We will see that, generally, these
constraints, together with the hot dark matter one, sufficiently restrict the 
range of family charges so that one tends to fall in the region of parameter space
favored by the LSND experiment.

\subsection{Atmospheric neutrino problem}
\indent

We start with the atmospheric neutrino problem which, if it remains, requires large
mixing angles, and thus in our framework, degenerate neutrinos.
In the model of Subsection $3.2$, the mass spectrum contains two heavy
quasi-degenerate neutrinos with a large mixing angle, and a light
neutrino which mixes weakly with the other ones ($m_{\nu_1} \ll -
m_{\nu_2} \simeq m_{\nu_3}$). In the model of Subsection 3.3
there are three quasi-degenerate neutrinos concomitant with large 
lepton mixing. Such patterns can account for the hot
dark matter of the Universe, and simultaneously explain the
atmospheric neutrino deficit in terms of $\nu_\mu - \nu_\tau$
oscillations \cite{2DEGNU,KAM,3DEGNU}.

  Consider first the model with two quasi-degenerate neutrinos. 
The parameters of the model are $l$ and $l'$ (with $0 < l < l'$), to 
which one should add $m_{\nu_3}$, whose value depends on the mass
scale of the heavy $RH$ neutrinos $\bar N_i$ involved in the seesaw.
Leaving out the arbitrary factors of order one and keeping only the
orders of magnitude, one can express all masses in units of
$m_{\nu_3}$ and in terms of $l$, $l'$: 
\begin{eqnarray}
  \frac{m_{\nu_1}}{m_{\nu_3}} \; \sim \; \epsilon^{\: 2l'},  &  &
  \frac{m_{\nu_2}}{m_{\nu_3}} \; \simeq \; - 1
\end{eqnarray}
\begin{eqnarray}
  \frac{\Delta m^2_{21}}{m_{\nu_3}^2} \simeq
	\frac{\Delta m^2_{31}}{m_{\nu_3}^2} \; \simeq \; 1,  &  &
  \frac{\Delta m^2_{32}}{m_{\nu_3}^2} \; \sim \; \epsilon^{\: 2l},
\end{eqnarray}
\begin{equation}
  \frac{|m_{\nu_{eff}}|}{m_{\nu_3}} \; \sim \; \epsilon^{\: 2l'}
  \label{eq:m_nu_eff}
\end{equation}
where\footnote{In our notations, $m_{\nu_i}$ is the mass of the $i^{th}$ Majorana 
neutrino ``signed'' by its $CP$-parity.} $m_{\nu_{eff}} = \sum_i m_{\nu_i} |U_{e i}|^2$ 
is the effective Majorana neutrino mass measured in neutrinoless double 
beta decay experiments. 
As for the mixing angles, their orders of magnitude are given by (\ref{eq:mixing}). 

When $CP$ violation in the lepton sector is neglected, 
the neutrino oscillation probability 
from one flavour $\alpha$ to another $\beta$ reads:
\begin{eqnarray}
P(\nu_\alpha \rightarrow \nu_\beta) & = & 
P(\bar{\nu}_\alpha \rightarrow \bar{\nu}_\beta)  =  \nonumber \\
      & & \delta_{\alpha \beta}
	- 4 \sum_{i<j} U_{\alpha i} U_{\beta i} U_{\alpha j} U_{\beta j} \,
	\sin^2 \left( \frac{\Delta m^2_{ji} L}{4 E} \right)
  \label{eq:oscillation}
\end{eqnarray}
where $E$ is the neutrino energy, $L$ the distance travelled by the neutrino 
between the source and the detector, and $U_{\alpha i}$ -- where the index 
$\alpha = e, \mu, \tau$ labels the weak eigenstates, while $i=1,2,3$ labels 
the mass eigenstates -- denote the entries of the lepton mixing matrix $U$. 
In the case of two heavy quasi-degenerate neutrinos, $\Delta m^2_{32} \ll 
\Delta m^2_{21} \simeq \Delta m^2_{31}$ and (\ref{eq:oscillation}) reduces to:
\begin{eqnarray} 
  P(\nu_\alpha \rightarrow \nu_\beta) & = & \delta_{\alpha \beta}
	- 4 U_{\alpha 1} U_{\beta 1} \left(\delta_{\alpha \beta} 
        - U_{\alpha 1} U_{\beta 1} \right)
	\sin^2 \left( \frac{\Delta m^2_{21} L}{4 E} \right)  \nonumber  \\
  & & - 4 \, U_{\alpha 2} U_{\beta 2} U_{\alpha 3} U_{\beta 3} \,
	\sin^2 \left( \frac{\Delta m^2_{32} L}{4 E} \right)
  \label{eq:oscillation2}
\end{eqnarray}
Due to the presence of two terms oscillating with very different frequencies 
($\Delta m^2_{32} \ll \Delta m^2_{21}$), one must distinguish between 
``short-distance'' and ``long-distance'' oscillations. Actually $P(\nu_\alpha 
\rightarrow \nu_\beta)$ depends on the ratio $L/E$, so ``short-distance'' 
regime means that $\frac{\Delta m^2_{32} L}{4 E} \ll 1$. This is the case 
for accelerator and reactor experiments. In this regime, $\nu_e$ - $\nu_\mu$, 
$\nu_\mu$ - $\nu_\tau$ and $\nu_e$ - $\nu_\tau$ oscillations 
are characterized by one and the same $\Delta m^2 \simeq \Delta m^2_{21}$, but 
with different oscillation amplitudes: 
\begin{eqnarray} 
  P_{sd}(\nu_\alpha \rightarrow \nu_\beta) & \simeq & 
	 4 U_{\alpha 1}^2 U_{\beta 1}^2 
         \sin^2 \left( \frac{\Delta m^2_{21} L}{4 E} \right)  \nonumber  \\
  & = & \sin^2 2 \theta_{\alpha\beta}^{sd} \,
	\sin^2 \left( \frac{\Delta m^2_{21} L}{4 E} \right), \, \alpha \neq \beta,
  \label{eq:oscillation3}
\end{eqnarray}
\noindent where in our scheme
\begin{equation}
\sin^2 2 \theta_{e \mu}^{sd}  \simeq  8  \, A^2  \,
		\epsilon^{\: 2l'-2l}, \,
\sin^2 2 \theta_{\mu \tau}^{sd} \simeq  8  \, A^2  \, B^2 \,
		\epsilon^{\: 4l'}, \, 
\sin^2 2 \theta_{e \tau}^{sd}  \simeq  4  \, B^2  \,
		\epsilon^{\: 2l'+2l} \, 
\end{equation}

\noindent However, if $\sin^2 2 \theta_{e \tau}^{sd}$ and 
$\sin^2 2 \theta_{\mu \tau}^{sd}$
are very small, as the above expressions suggest, 
the corrections in the probabilities
$P_{sd}(\nu_e \rightarrow \nu_\tau)$ and $P_{sd}(\nu_\mu \rightarrow \nu_\tau)$
due to the long wave length oscillation term in eq. (45) can be important: 
they are given respectively by the terms  
$2\left( \frac{\Delta m^2_{32} L}{4 E} \right)^2 \, A^2 \,  
		\epsilon^{\, 2l'-2l}$ and  
$\left( \frac{\Delta m^2_{32} L}{4 E} \right)^2$. Actually, as it follows
from eq. (46), the expression for the relevant correction and the fact
that $l' > l$, the ``long distance'' term is always dominant in 
$P_{sd}(\nu_\mu \rightarrow \nu_\tau)$.

In the "long-distance" regime $\frac{\Delta m^2_{21} L}{4 E} \gg 1$ and 
the oscillations corresponding to
the first term in (\ref{eq:oscillation2}) are averaged out. 
The probabilities relevant for atmospheric neutrinos are:
\begin{eqnarray} 
  P_{ld}(\nu_\mu \rightarrow \nu_e) & \simeq &
	4 \, A^2 \, \epsilon^{\: 2l'-2l} \:
	\left[ 1 - \frac{1}{2} \,
	\sin^2 \left( \frac{\Delta m^2_{32} L}{4 E} \right) \right]
\end{eqnarray}
\begin{eqnarray} 
  P_{ld}(\nu_\mu \rightarrow \nu_\tau) & \simeq &
	\sin^2 \left( \frac{\Delta m^2_{32} L}{4 E} \right) 
       + 4 \, A^2 \, B^2 \, \epsilon^{\: 4l'} \:  
\end{eqnarray}
Clearly $\nu_\mu$ - $\nu_e$ oscillations are suppressed compared with 
$\nu_\mu$ - $\nu_\tau$ oscillations. 

Let us now show that the model considered can account for the 
atmospheric neutrino anomaly and the dark matter problem
simultaneously. The requirement that the mu and the tau neutrinos
constitute the hot dark matter fixes the mass scale $m_{\nu_3}$ to be
in the few $eV$ range, typically $m_{\nu_3} = 2-3 \: eV$
\cite{Primack}. The solution of the atmospheric neutrino problem in
terms of $\nu_\mu$ - $\nu_\tau$ oscillations requires  \cite{KAM}
$5.10^{-3} \: eV^2 \leq \Delta m^2_{32} \leq 3.10^{-2} \: eV^2$. We must 
also take into account the experimental limits on
$\nu_e$ - $\nu_\mu$ oscillations. In the few $eV^2$
region, the experimental upper bound on $\sin^2 2 \theta^{sd}_{e \mu}$ from 
E776 is \cite{E776} $(2-3).10^{-3}$. 
%while the LSND result requires
%\cite{LSND} $2.10^{-3} \leq \sin^2 2 \theta_{e \mu} \leq 10^{-2}$. 
These data strongly constrain the parameters of the model. 
Assuming that $\epsilon$ is the Cabbibo angle ($\epsilon \simeq 0.22$), 
as suggested by the quark sector, we obtain:
 \begin{eqnarray}
  l = 2 \pm 1/2  &  \mbox{and}  &  l'-l \geq 3
\end{eqnarray}
With such values of $l$ and $l'$, only a few charge assignments for the charged 
lepton singlets $\bar E_i$ reproduce the observed mass hierarchy in the charged 
lepton sector. Indeed, under our hypotheses, we have
\begin{equation}
{m_e \over m_\tau} = \epsilon^{l'+l+e_1 -e_3}.
\end{equation}
For example, the following choice:
\begin{eqnarray}
  \left \{  \begin{array}{l}
	l' = 5  \\
	l = 2
  \end{array}  \right.  &  \mbox{and}  &
  \left \{  \begin{array}{l}
	e_1 = e_2 = e_3 - 2  \\
	2 \leq e_3 < 4
  \end{array}  \right.
\end{eqnarray}
leads to $m_\mu / m_\tau \sim \epsilon^{\: 2}$ and $m_e / m_\tau \sim \epsilon^{\: 5}$ 
(which deviates slightly from the geometrical hierarchy $1 : \epsilon^{\: 2} : \epsilon^{\: 4}$). 
The features of the model are then:

\begin{enumerate}
	\item the neutrino mass spectrum contains two quasi-degenerate 
neutrinos in the few $eV$ range, which can account for the hot dark
matter: \begin{eqnarray}
  m_{\nu_e} \simeq m_{\nu_1} \sim (5-8).10^{-7} \: eV  &  &
	m_{\nu_\mu} \simeq m_{\nu_\tau} \simeq m_{\nu_3} = 2-3 \: eV
\end{eqnarray}
	\item the atmospheric neutrino anomaly is explained by 
$\nu_\mu - \nu_\tau$ oscillations with parameters:
\begin{eqnarray}
  \Delta m^2_{\mu \tau} \simeq \Delta m^2_{32} \sim (0.9-2.1).10^{-2} \: 
eV^2  &  \mbox{and}  &
	\sin^2 2 \theta_{\mu \tau} \simeq 1
\end{eqnarray}
	\item one finds that $\bar \nu_\mu - \bar \nu_e$ oscillations may 
be compatible with
the positive result announced by the LSND collaboration and can be
observed by KARMEN \cite{KARMEN}: 
\begin{eqnarray}
  \Delta m^2_{e \mu} \simeq \Delta m^2_{21} = 4-9 \: eV^2  &  \mbox{and}  &
	\sin^2 2 \theta_{e \mu} \simeq 0.9 A^2 \, 10^{-3}
\end{eqnarray}
with $0.2 < A < 5$. 
	\item $\nu_e - \nu_\tau$ and $\nu_\mu - \nu_\tau$ oscillations are below 
the sensitivity of CHORUS 
and NOMAD \cite{NOMAD}, but  $\nu_\mu - \nu_\tau$ oscillations may be observable in 
future long-baseline experiments. 
	\item neutrinoless double beta decay rate is far below the sensitivity 
of current and planned experiments (see, e.g., \cite{HM}):
	\begin{equation}
	  |m_{\nu_{eff}}| \; \ll 10^{-2} \; eV
	\end{equation}
 the solar neutrino problem cannot be solved
by this model, unless a sterile neutrino $\nu_s$ is added \cite{4NU}, 
since the
squared mass difference required by the MSW solution ($\Delta m^2 \sim
10^{-6}-10^{-5} \: eV^2$) or the vacuum oscillation solution ($\Delta
m^2 \sim 10^{-11}-10^{-10} \: eV^2$) \cite{KP} is much less than $\Delta
m^2_{32}$ and $\Delta m^2_{21}$.

\end{enumerate}

  Let us note that the neutrino phenomenology of such a model is 
somewhat close to the one
encountered in the Zee model \cite{Zee}, as discussed recently by Smirnov and
Tanimoto \cite{ST} (see also \cite{LWSP}). Although the models and their 
physical motivations are very different, it
would not be so easy to distinguish between the two, except if a signal for 
$\nu_e - \nu_\tau$ oscillations is found in CHORUS and NOMAD.

 Finally, let us comment briefly on the solution of the atmospheric neutrino problem
in the model of Subsection 3.3 with three quasi-degenerate neutrinos. In this case there 
are essentially 
two parameters: the mass $|m_{\nu_{2}}| \simeq a \equiv m > 0$ and the charge $l$. 
One has to leading order in $\epsilon^{\: l}$:
\begin{equation}
  m_{\nu_1} \; \simeq |m_{\nu_2}| \; \simeq \; m_{\nu_3} \; \simeq \; m,
  \label{eq:3degn1}
\end{equation}
\begin{equation}
  \Delta m^2_{31} \; \simeq \;
	2\Delta m^2_{21} \; \simeq \;
  2\Delta m^2_{32} \; \simeq \; 2 m^2 \epsilon^{\: l},
  \label{eq:3degn2}
\end{equation}
\begin{equation}
  |m_{\nu_{eff}}| \; \simeq \; m,
  \label{eq:m_nu_eff2}
\end{equation}
\noindent where we have used $(c + e) \sim a$ in obtaining eq. (60).

The massive neutrinos can constitute the hot component in the two component 
``hot + cold'' dark matter theory provided $m \simeq (1.5 - 2.0)~eV$.  
This implies in our model $|m_{\nu_{eff}}| \simeq (1.5 - 2.0)~eV$, which is 
by approximately a factor of 2 larger than the most stringent
upper limit on $|m_{\nu_{eff}}|$, quoted in the literature on the subject
and derived from the negative results of 
the searches for neutrinoless double-beta decay of $^{76}$Ge \cite{HVK}.
However, in view of the uncertainties in the calculations of the nuclear matrix elements
entering into the expression for the neutrinoless double-beta decay amplitude,
values of $|m_{\nu_{eff}}| \simeq 1.5~eV$ cannot be ruled out
by the presently existing data (see, e.g., \cite{PV96}). Ongoing and future 
experiments will test relatively soon this possibility 
\footnote{The neutrinoless 
double-beta decay rate will be suppressed if the quasi-degeneracy of the masses of the
three neutrinos is a result of the conservation of the charge 
$X_{L} = l\, (L_{\mu} - L_e)$ (instead of the charge $X_{L} = l\, (L_{\mu} - L_{\tau})$ 
in the example we are considering) before the breaking of the $U(1)_{X}$ symmetry.}.

The atmospheric neutrino problem can be solved in the model under consideration in terms
of three-neutrino oscillations of the $\nu_{\mu}$ and $\bar{\nu}_{\mu}$, characterized
by two $\Delta m^2$ which differ by a factor of two. A three-neutrino oscillation
analysis of the Kamiokande data suggests \cite{OY96} 
$\Delta m^2_{31} \; \simeq \; 2\Delta m^2_{21} \; \simeq (2 - 4).10^{-2}~eV^2$.
With $m$ in the cosmologically relevant region this implies $l = 3~ or~4$. 
The  $\nu_{\mu} - \nu_{e}$  and $\nu_{\mu} - \nu_{\tau}$ oscillation 
probabilities have a very simple form: 
\begin{eqnarray} 
 & & P(\nu_\mu \rightarrow \nu_e)  \simeq \frac{1}{2} \,
	\sin^2 \left( \frac{2\Delta m^2_{21} L}{4 E} \right), \nonumber \\
 & & P(\nu_\mu \rightarrow \nu_\tau) \simeq
	\sin^4 \left( \frac{\Delta m^2_{21} L}{4 E} \right)   
\end{eqnarray}
\noindent Note that the amplitude of the $\nu_{\mu} - \nu_{\tau}$ oscillations
is larger than the amplitude of the  $\nu_{\mu} - \nu_{e}$ oscillations.
 
The model predicts neutrino energy-independent suppression of the different
components of the solar neutrino flux by the factor 0.5. Such a suppression 
is not favoured by the current solar neutrino data \cite{KP4}.

\subsection{Adiabatic MSW effect}
\indent

We now address the possibility that the model of Subsection $3.2$ 
be used to solve the solar neutrino problem along the lines of the MSW interpretation in 
the large angle branch. The scale pattern is then: 
$m_{\nu_\mu} \ll m_{\nu_e} \simeq m_{\nu_\tau}$.
Due to the inverted hierarchy in the electron-muon 
sector, the model 
has to be slightly modified. The relevant charge assignment is now:
\begin{eqnarray}
  \left\{ \begin{array}{l}
	l_2 = l' \\ l_1 = - l_3 = l \\ 0 < l' < l  \end{array}  \right.
  &  \mbox{and}  &  0 \leq n_1, l' < (-n_3) \leq l \leq n_2
\end{eqnarray}
corresponding to the following pattern for $\widehat{{\cal M}}_\nu$:
\begin{equation}
  \widehat{\cal M}_\nu \simeq \left( \begin{array}{ccc}
    \epsilon^{\: 2l} &  \epsilon^{\: l'+l} & 1 \\
    \epsilon^{\: l'+l} & \epsilon^{\: 2l'} &
	\epsilon^{\: l-l'} \\
    1 & \epsilon^{\: l-l'}  &
	 \epsilon^{\: 2l}   \end{array}  \right) \label{eq:hatMnu'}
\end{equation}
where again only the dominant term in each entry is given. This yields mass ratios:
\begin{eqnarray}
  \frac{m_{\nu_2}}{m_{\nu_3}} \; \sim \; \epsilon^{\: 2l'}  &  &
  \frac{m_{\nu_1}}{m_{\nu_3}} \; \simeq \; - 1
\end{eqnarray}
\begin{equation}
  \frac{\Delta m^2_{31}}{m_{\nu_3}^2} \; \sim \; \epsilon^{\: 2l}
\end{equation}

The requirement that the model account for the solar neutrino deficit 
by adiabatic MSW transitions and solve the dark matter problem simultaneously, 
together with the experimental constraints on $\nu_e - \nu_\mu$ oscillations and 
the measured values of 
the charged lepton masses, leads to the following choice of parameters:
\begin{eqnarray}
  \left \{  \begin{array}{l}
	l' = 3/2  \\
	l = 9/2
  \end{array}  \right.  &  \mbox{and}  &
  \left \{  \begin{array}{l}
	e_1 = e_2 = e_3 - 4  \\
	9/2 \leq e_3 < 17/2
  \end{array}  \right.
\end{eqnarray}
Again, the mass hierarchy in the charged lepton sector 
slightly departs from the geometrical 
scheme: $\frac{m_\mu}{m_\tau} \sim \epsilon^{\: 2}$ and 
$\frac{m_e}{m_\tau} \sim \epsilon^{\: 5}$. 
The features of the model are:
\begin{enumerate}
	\item the electron and the tau neutrinos are 
quasi-degenerate in mass and can account for the hot dark matter:
\begin{eqnarray}
  m_{\nu_\mu} \simeq m_{\nu_2} \sim 0.03 \: eV  &  &
	m_{\nu_e} \simeq m_{\nu_\tau} \simeq m_{\nu_3} = 2-3 \: eV
\end{eqnarray}
	\item the solar neutrino deficit is explained by adiabatic 
MSW $\nu_e - \nu_\tau$ transitions with parameters:
\begin{eqnarray}
  \Delta m^2_{e \tau} \simeq \Delta m^2_{31} \sim (0.5-1.1).10^{-5} \: eV^2  
&  \mbox{and}  &
	\sin^2 2 \theta_{e \tau} \simeq 1
\end{eqnarray}
	\item $\bar \nu_\mu - \bar \nu_e$ oscillations may be in 
the domain of sensitivity of LSND and KARMEN: 
\begin{eqnarray}
  \Delta m^2_{e \mu} \simeq \Delta m^2_{32} = 4-9 \: eV^2  &  
\mbox{and}  &
	\sin^2 2 \theta_{e \mu} \simeq A^2 \, 10^{-3}
\end{eqnarray}
with $0.2 < A < 5$.
	\item $\nu_e - \nu_\tau$ and $\nu_\mu - \nu_\tau$ oscillations are below 
the sensitivity of current and 
planned experiments:
\begin{equation}
	\left \{ \begin{array}{l}
  P_{sd}(\nu_e \rightarrow \nu_\tau) \; \simeq \;
	\sin^2 (\Delta m^2_{e \tau} L/4E)  \\
  P_{sd}(\nu_\mu \rightarrow \nu_\tau) \; \leq \; 5.10^{-8}
	\end{array}  \right.
\end{equation}
        \item non-leading terms in the lepton mixing matrix can yield a contribution to
$ |m_{\nu_{eff}}|$ of the order of $m_{\nu_3} \epsilon^{2l'}$, in the border range of 
sensitivity of planned neutrinoless double-beta decay
experiments.
\end{enumerate}
Note, however, that the solar neutrino data tends to favour smaller 
mixings between the electron and the tau neutrino than predicted by the model, 
namely $0.2 \leq \sin^2 2 \theta_{e \mu} \leq 0.9$.

   In the model with three quasi-degenerate neutrinos of Subsection 3.3 the solar 
neutrino problem can be solved in terms of three-neutrino MSW transitions.
Neutrinos can have masses in the cosmologically relevant range of (1.5 - 2.0) eV
and in this case the neutrinoless double-beta decay rate may not be suppressed.  
Obviously, this model cannot provide an explanation of the atmospheric neutrino anomaly 
and/or of the LSND result if $\Delta m^2_{21}$ (and therefore $\Delta m^2_{31}$)
has a value required by the MSW solution of the solar neutrino problem.

\section{Conclusions}

It certainly follows a natural path to try and apply the recent ideas 
on family symmetries -- developped to explain the observed hierarchies of 
masses and mixings in the sector of quarks and charged leptons -- to the
poorly known neutrino sector. There, the promises of new experimental results
in the near future
make it a natural ground for theorists to make predictions. Also the models considered
put the problem of neutrino masses in the perspective of a more general framework
which deals with the masses and mixings of all the low-energy particles.

We have considered in this paper abelian family gauge symmetries,  
because of both their simplicity and their attractive features in the quark sector.
In the case where the third family, including the superheavy right-handed neutrino,
is the heaviest of the three, one obtains predictions in the low-energy 
neutrino sector which strikingly ressemble those of the quark sector. 
In particular, neutrino masses have a hierarchical structure and,
much like the CKM matrix, the neutrino mixing angles can be, 
to a first approximation, expressed as ratios of neutrino masses 
\cite{DLLRS,GN,Montreal,BLR}. We tend to favor such a case, although it 
cannot be reconciled, as is, with {\em all} the present neutrino experimental 
results --in particular because the mixing angles are small.

In this paper, we have on the other hand considered the situation where
one, or more, mixing angle are large. We have showed that this is
perfectly compatible with an abelian gauge family symmetry. 
The spectrum of light neutrinos then involves some level of degeneracy.
The models in this class are rather constrained. They can be used to address more particularly 
the atmospheric neutrino problem or the solar neutrino problem in its large 
angle MSW interpretation. We have studied the phenomenological consequences 
of such models for present and foreseen neutrino experiments and conclude that 
they are within the range of such experiments. 

\vskip 1cm
{\bf Acknowledgments.}
\vskip .5cm

P.B. and P.R. would like to thank the Aspen Center for Physics where part of this work
was completed. They also acknowledge a CNRS-NSF fund INT-9512897 for this research.
The work of S.P. was supported in part by the EEC grant ERBCHRX CT930132 and by the
grant PH-510 from the Bulgarian Science Foundation.

%%%%%%%%%%%%%%%%%%%%%%%%%%%%%%%%%%%%%%%%%%%%%%%%%%%%%%%%%%%%%%%%%%%

%%%%%%%%%%%%%%%%%%%%%%%%%%%%%%%%%%%%%%%%%%%%%%%%%%%%%%%%%%%%%%


\begin{thebibliography}{Ref}
\bibitem{FN} C.~Froggatt and H.~B.~Nielsen Nucl. Phys. B147 (1979)
277. 

\bibitem{LNS} M. Leurer, Y. Nir, and N. Seiberg, Nucl. Phys. B398
(1993) 319, Nucl. Phys. B420 (1994) 468. 

\bibitem{IR} L. Ib\'a\~nez
and G. G. Ross, Phys. Lett. B332 (1994) 100. 

\bibitem{BR} P. Bin\'etruy and P. Ramond, Phys. Lett. B350 (1995) 49. 

\bibitem{DLLRS} H. Dreiner, G.K. Leontaris, S. Lola, G.G. Ross and C.
Scheich, Nucl. Phys. B436 (1995) 461; G.K. Leontaris, S. Lola, C.
Scheich and J.D. Vergados, Phys. Rev. D53 (1996) 6381. 

\bibitem{Elena} E. Papageorgiu, Z. Phys. C64 (1994) 509, Phys. Lett. B343 (1995) 263, 
hep-ph/9510352.

\bibitem{GN} Y. Grossman and Y. Nir, Nucl. Phys. B448 (1995) 30.

\bibitem{Montreal} P. Ramond, 25th Anniversary Volume of the Centre de
Recherches Math\'ematiques de l'Universit\'e de Montr\'eal, hep-ph/9506319.

\bibitem{BLR} P. Bin\'{e}truy, S. Lavignac and P. Ramond, 
Nucl. Phys. B477 (1996) 353.

\bibitem{MSW}
S.P. Mikheyev and A.Yu. Smirnov,
Yad. Fiz. {\bf 42}, 1441 (1985)
[Sov. J. Nucl. Phys. {\bf 42}, 913 (1985)];
Il Nuovo Cimento C {\bf 9}, 17 (1986);
L. Wolfenstein,
Phys. Rev. D {\bf 17}, 2369 (1978);
Phys. Rev. D {\bf 20}, 2634 (1979).

\bibitem{2DEGNU} S.T. Petcov and A.Yu. Smirnov, Phys. Lett. B322 (1994) 109;
D.O. Caldwell and R.N. Mohapatra, Phys. Lett. B354 (1995) 371;
G. Raffelt and J. Silk, Phys. Lett. B366 (1996) 429.

\bibitem{2DEGNUPH} S.M. Bilenky et al., Phys. Rev. D54 (1996) 4432.

\bibitem{3DEGNU} D.O. Caldwell and R.N. Mohapatra, Phys. Rev. D48 (1993) 
3259 and D50 (1994) 3477.
% Z. Berezhiani and R.N. Mohapatra, Phys. Rev. D52 (1995) 6607.

\bibitem{seesaw} M. Gell-Mann, P. Ramond, and R. Slansky in Sanibel 
Talk, CALT-68-709, Feb 1979, and in {\it Supergravity} (North Holland,
Amsterdam 1979). T. Yanagida, in {\it Proceedings of the Workshop on
Unified Theory and Baryon Number of the Universe}, KEK, Japan, 1979.

\bibitem{JS} V. Jain and R. Shrock, Phys. Lett. B352 (1995) 83. 

\bibitem{DPS} E. Dudas, S. Pokorski and C.A. Savoy, Phys. Lett. B356
(1995) 45.

\bibitem{Nir} Y. Nir, Phys. Lett. B354 (1995) 107.

\bibitem{Ibanez} L. Ib\'a\~nez, Phys. Lett. B303 (1993) 55. 

\bibitem{DSW} M. Dine, N. Seiberg and E. Witten, Nucl. Phys. B289
(1987) 589; J. Atick, L. Dixon and A. Sen, Nucl. Phys. B292 (1987)
109; M. Dine, I. Ichinose and N. Seiberg, Nucl. Phys. B293 (1988) 253.

\bibitem{Pont57} B. Pontecorvo, Zh. Eksp. Teor. Fiz. 33 (1957) 549 and 34 (1958) 247.

\bibitem{MNS62} Z. Maki, M. Nakagawa and S. Sakata, Prog. Theor. Phys. 28 (1962) 870.

\bibitem{GP69} V. Gribov and B. Pontecorvo, Phys. Lett. 28B (1969) 493. 

\bibitem{P} S.T. Petcov, Phys. Lett. 110B (1982) 245; C.N. Leung and S.T.
Petcov, Phys. Lett. 125B (1983) 461.

\bibitem{WW} D. Wyler and L. Wolfenstein, Nucl. Phys. B218 (1983) 205.

\bibitem{CP} L. Wolfenstein, Phys. Lett. B107 (1981) 77; S.M.
Bilenky, N.P. Nedelcheva and S.T. Petcov, Nucl. Phys. B247 (1984) 61;
B. Kayser, Phys. Rev. D30  (1984) 1023.

\bibitem{ZKM} Ya.B. Zeldovich, DAN SSSR 86 (1952) 505;
E.J. Konopinsky and H. Mahmoud, Phys. Rev. 92 (1953) 1045.

\bibitem{SNP}
% R. Davis, Prog. Part. Nucl. Phys. {\bf 32}, 13 (1994); 
B.T. Cleveland et al., Nucl. Phys. B (Proc. Suppl.) 38 (1995) 47;
% \bibitem{Kamiokande}
Y. Suzuki (Kamiokande Coll.), talk given at the ``Neutrino '96'' Int. Conf., 
June 13 - 19, Helsinki, Finland;
% K. S. Hirata et al., Phys. Rev. D44 (1991) 2241;
% \bibitem{GALLEX}
GALLEX Coll.,
Phys. Lett. B357 (1995) 237;
% \bibitem{SAGE}
V. Vermul (SAGE Coll.),
Talk presented at the
{\it $7^{th}$ International Workshop
on Neutrino Telescopes},
February 1996, Venice;
% \bibitem{bahcall}
% J.N. Bahcall and R. Ulrich,
% Rev. Mod. Phys. {\bf 60}, 297 (1988);
J.N. Bahcall,
{\it Neutrino Physics and Astrophysics},
Cambridge University Press, 1989;
J.N. Bahcall and M.H. Pinsonneault,
% Rev. Mod. Phys. {\bf 64}, 885 (1992) and
Rev. Mod. Phys. 67 (1995) 1.

\bibitem{VOS} B. Pontecorvo, JETP 53 (1967) 1717;
S.M. Bilenky and B. Pontecorvo, Phys. Rep. 41 (1978) 225.

\bibitem{KP} P.I. Krastev and S.T.
Petcov, Nucl. Phys. B449 (1995) 605.

\bibitem{KAM} Y. Fukuda et al., Phys. Lett. B335 (1994) 237.

\bibitem{ANP1} R. Becker-Szendy et al., Nucl. Phys. B (Proc. Suppl.) 38 
(1995) 
331; M. Goodman et al., Nucl. Phys. B (Proc. Suppl.) 38 (1995) 337.

\bibitem{ANP2} T.K. Gaisser, F. Halzen and T. Stanev,
Phys. Rep. 258 (1995) 173; see also:
W. Frati, T.K. Gaisser, A.K. Mann and T. Stanev,
Phys. Rev. D48 (1993) 1140.

\bibitem{LSND} C. Athanassopoulos
et al., Phys. Rev. Lett. 75 (1995) 2650, and Los Alamos Nat.
Lab. report LA--UR--96--1326, April 24, 1996; see  also J.E. Hill,
Phys. Rev. Lett. 75 (1995) 2654.  

\bibitem{Primack} J.R. Primack et al., Phys. Rev. Lett. 74 
(1995) 2160. 

\bibitem{E776} L. Borodovsky et al., Phys. Rev. Lett. 68 (1992) 274.

\bibitem{KARMEN} KARMEN Collaboration, Nucl. Phys. B38 (Proc. Suppl.)
(1995) 235. 

\bibitem{NOMAD} K. Winter, Nucl. Phys. B38 (Proc. Suppl.)
(1995) 211; M. Baldo-Ceolin, Nucl. Phys. B35 (Proc. Suppl.) (1994)
450; L. DiLella, Nucl. Phys. B31 (Proc. Suppl.) (1993) 319.

\bibitem{HM}  M. Moe and P. Vogel, Annu. Rev. Nucl. Part. Sci. 44 (1994) 247;
M.K. Moe,  Nucl. Phys. B (Proc. Suppl.) 38 (1995) 36; see also: A. Balysh et al.,
Phys. Lett. B356 (1995) 450.

\bibitem{4NU} J.T. Peltoniemi and J.W.F. Valle,
Nucl. Phys. B406 (1993) 409; Z. Shi, D.N. Schramm and B.D.
Fields, Phys. Rev. D48 (1993) 2563; J.J. Gomez-Cadenas and
M.C. Gonzalez-Garcia, preprint CERN-TH/95-80, hep-ph/9504246; M.C.
Gonzalez-Garcia, preprint CERN-TH/95-285; hep-ph/9510419; S. Goswami,
preprint CUPP-95/4, hep-ph/9507212. 

\bibitem{Zee} A. Zee, Phys. Lett. 93B (1980) 389, Phys. Lett. 161B (1985) 141.

\bibitem{ST} A. Smirnov and M. Tanimoto, preprint EHU-96-4 (April 1996).

\bibitem{LWSP} L. Wolfenstein, Nucl. Phys. B175 (1980) 93;
S.T. Petcov, Phys. Lett. 115B (1982) 401.

\bibitem{HVK} M. Gunther et al. (Heidelberg-Moscow coll.), Proc. of the Int. Workshop 
          on Double--Beta Decay and
          Related Topics, April 24 -- May 5, 1995, European Centre for 
          Theoretical Studies, Trento, Italy (eds. H.V. Klapdor-Kleingrothaus and
          S. Stoica, World Scientific, Singapore, 1996), p. 455.

\bibitem{PV96} P. Vogel, Proc. of the Int. Workshop on Double--Beta Decay and
          Related Topics, April 24 -- May 5, 1995, European Centre for 
          Theoretical Studies, Trento, Italy (eds. H.V. Klapdor-Kleingrothaus and
          S. Stoica, World Scientific, Singapore, 1996), p. 323.

\bibitem{OY96} O. Yasuda, Tokyo Metropolitan Univ. Report TMUP-HEL-9603, February 1996.

\bibitem{KP4} P.I. Krastev and S.T. Petcov, Phys. Rev. D53 (1996) 1665.

\end{thebibliography}
\end{document}